\documentclass[12pt,preprint]{aastex}

\usepackage{amsmath}
\usepackage[colorlinks,linkcolor=red,anchorcolor=black,citecolor=blue]{hyperref}

\shorttitle{Anisotropy as a Probe of the GCR Propagation and GMF}

\shortauthors{Xiao-bo Qu et al.}

\begin{document}

\title{Anisotropy as a Probe of the Galactic Cosmic Ray Propagation and Halo Magnetic Field}

\author{Xiao-bo Qu\altaffilmark{1}, Yi Zhang\altaffilmark{1,*}, Liang Xue\altaffilmark{2}, Cheng Liu\altaffilmark{1},  Hong-bo Hu\altaffilmark{1}}
\affil{$^1$Key Laboratory of Particle Astrophysics, Institute of High Energy Physics,
    Chinese Academy of Sciences, Beijing 100049, China}

\affil{$^2$School of Physics, Shandong University, Ji'nan 250100, China}
%\affil{$^\ast$To whom correspondence should be addressed; E-mail: zhangyi@mail.ihep.ac.cn}
\altaffiltext{*}{corresponding author: zhangyi@mail.ihep.ac.cn}

\begin{abstract}
 The anisotropy of cosmic rays (CRs) in the solar vicinity is generally attributed to the CR streaming due to the discrete distribution of CR sources or local magnetic field modulation.  Recently, the two dimensional large scale CR anisotropy has been measured by many experiments in TeV-PeV energy range in both hemispheres. The tail-in excess along the tangential direction of the local spiral arm and the loss cone deficit pointing to the north Galactic pole direction agree with what have been obtained in tens to hundreds of GeV. The persistence of the two large scale anisotropy structures in such a wide range of energy suggests that the anisotropy might be due to a global streaming of the Galactic CRs (GCRs). This work tries to extend the observed CR anisotropy picture from solar system to the whole galaxy. In such a case, we can find a new interesting signature, a loop of GCR streaming, of the GCR propagation. We further calculate the overall GCR streaming induced magnetic field, and find a qualitative consistence with the observed structure of the halo magnetic field.
\end{abstract}


\section{INTRODUCTION}

Galactic magnetic field (GMF), Galactic cosmic ray (GCR) and the ordinary matter are the basic components of the interstellar medium (ISM). These three constituents have comparable pressure and are bounded together by the electromagnetic force. The GMF and the propagation of GCRs help to support the ordinary matter against the self-gravity. Conversely, the weight of the ordinary matter confines GMF and then GCRs in the Galaxy \citep{fer01}. The dynamic balance process between these three constituents could give birth to new molecular-cloud complexes and ultimately trigger star formation \citep{mou74,elm82}. The physical nature of and the interactions between them are studied for decades. However, the fundamental question, the origin of them, is still open. Nowadays, due to the development of the observation technology and methods, the information about these constituents is more adequate, and we should approach a better understanding of this question.

The GMF is commonly believed to be produced by a dynamo process with a seed field (e.g., see review \citep{bra05} and references therein). Based on the measurements of the Faraday rotation of the linearly polarized radiation from pulsars and extragalactic radio sources \citep{sim81,con98,bro03,han09,tay09,law11}, two kinds of structures are found in the halo magnetic field: the poloidal fields of dipole structure perpendicular to the plane in the Galactic center (GC) region and the toroidal fields of opposite directions above and below the Galactic plane \citep{han97,han99}. The \textquotedblleft A0\textquotedblright dynamo model \citep{han97} could produce such a large scale magnetic field. Nevertheless, the origin of the GMF is still an open question.

The GCR streaming, i.e. GCR propagation, could have contributions to the GMF. \citet{par92,hana09} suggested that the GMF could be generated by the dynamo process driven by GCR streaming. \citet{dol04} proposed a mechanism to produce the large-scale GMF by the electric current induced by GCRs. In their model, the current was theoretically calculated based on the diffusion law, and the result depends strongly on the assumption of the diffusion coefficient tensor. However, there is no direct measurement of the diffusion coefficient. It is better to use a direct method to estimate the GCR streaming.

The anisotropy of GCRs should be a precise tool to probe the GCR streaming. Early in 1935, \citet{com35} suggested that the relative movement between the GCR plasma and the observer due to the Galactic rotation could lead to a dipole anisotropy, known as the Compton-Getting (CG) effect. According to this effect, the direction and velocity of the CR streaming can be estimated by the phase and amplitude of the anisotropy respectively. \citet{ame06} did precise measurement above 300 TeV and concluded that the GCR plasma might corotate with the solar environment around the Galactic center with a velocity ${\sim}220 \ km s^{-1}$. In such a manner, the CR streaming in the rest frame of the Galaxy could be deduced from the high-precision two-dimensional anisotropy map, which is observed by many experiments in a wide energy range.

In this work, we attempt to extend the anisotropy pattern observed in the solar vicinity to the whole Galaxy. The GCR streaming related to the anisotropy should exist in Galactic scale, forming a new picture of the GCR propagation. Furthermore, the global streaming of the GCR can be used to explore the contribution of GCR to the GMF. This Letter is organized as follows: In Section 2, we introduce the observational anisotropy and the GCR streaming in the extended picture. In Section 3, we estimate the contribution of the GCR to the GMF in the halo. In Section 4, we present discussions and the conclusions of this extension.

\section{ANISOTROPY AND COSMIC RAY PROPAGATION}

There is a long history to study the GCR anisotropy, with both underground $\mu$ detectors and ground-based air shower arrays \citep{agl96,mun97,amb03,ant04,ame05}. Owing to the long-term observations and newly developed analysis methods, the high-precision two-dimensional map of the GCR anisotropy was obtained by Tibet air shower array \citep{ame06}, Super-Kamiokande \citep{gui07}, MILAGRO \citep{abd09}, ARGO \citep{zha09} in the northern hemisphere, and IceCube \citep{abb10} in the southern hemisphere. In this two-dimensional map, three broad structures are distinctly presented: an excess called the \textquotedblleft tail-in\textquotedblright, a deficit called \textquotedblleft loss-cone\textquotedblright and an excess in the direction of the Cygnus region.

Figure 1 (a) shows a schematic view of the GCR streamings in the solar system based on the northern hemisphere observations of the anisotropy. The three broad structures are shown as follows.

The center of the \textquotedblleft tail-in\textquotedblright \ component is at ($Dec {\sim}{-22}^{\circ}30', R.A. {\sim} 97^{\circ}24'$), close to the direction of the heliomagnetic tail ($Dec {\sim}{-29}^{\circ}12', R.A. {\sim} 90^{\circ}24'$) which is opposite to the proper motion direction of the solar system \citep{dul01}. This direction is quite close to the tangential direction of the local arm, indicating an inward GCR streaming along the local arm. A slight difference between the directions of the GCR streaming and the local arm might be due to the deflection by the heliomagnetic field.

The center of the \textquotedblleft loss-cone\textquotedblright \ deficit component points to the direction of the north Galactic pole (NGP). We interpret it as an outward streaming toward the NGP, the direction of which is perpendicular to the Galactic plane. In this picture, the GCR anisotropy of the southern hemisphere should show an excess toward the direction of the south Galactic pole (SGP) ($Dec {\sim}{-27}^{\circ}6', R.A. {\sim} 12^{\circ}48'$). The observation from IceCube indeed showed such an excess  \citep{abb10}.

The excess component related to the Cygnus region peaks at ($Dec {\sim}38^{\circ}, \\ R.A. {\sim} 309^{\circ}$). Similar with the above discussion, we expect this component to indicate an outward GCR streaming along the local spiral arm.

The structures of the \textquotedblleft tail-in\textquotedblright \ and \textquotedblleft loss-cone\textquotedblright \ components are almost stable in the energy range between tens of GeV and hundreds of TeV, and insensitive to the solar activities \citep{ame10}. This indicates that the GCR anisotropy in this energy range may not be related to the heliospheric magnetic field, but of a large-scale origin. There is no reason that our solar location is special in the Galaxy. Therefore, we can extend the above-mentioned anisotropy pattern to the whole Galaxy. After that, the GCR plasma should have global streamings in three directions: inward and outward streamings along the spiral arms (The inward streaming related to the \textquotedblleft tail-in\textquotedblright \ region is significantly larger than the outward streaming related to the Cygnus region. Thus, the overall streaming along the local arm is inward), and outward streaming perpendicular to the Galactic disk, as shown in Figure 1 (b).

This global picture of the GCR propagation could be understood in the following way. The diffusion of GCRs is constrained by the magnetic field line, and the direction of the GCR streaming is decided by the gradient of the GCR intensity. Since most of the GCR sources are located in the inner Galactic disk, GCRs tend to diffuse out of the Galaxy in a direction perpendicular to the Galactic plane. There is a streaming back into the Galaxy from the direction of the spiral arm. Such a streaming may compensate the overall charge loss of the outward streaming, and makes the loop streaming of GCRs possible. Such a global loop streaming is not a new concept in astrophysics. For example, \citet{alf78} has discussed the electric double layers phenomenon and related loop electric current from laboratory scale to terrestrial, stellar and galactic scale.

The streaming of the GCR plasma can form an electric current. The direction of the current is the same as the direction of the streaming because the GCR particles are mainly protons with positive charge (the fraction of relativistic electrons is less than 1\%). The electric current density can be calculated by

\begin{equation}
j_{cr} = q\int \frac{dN(E)}{dE}v(E)dE.
\end{equation}
Here $q$ is the mean charge of the GCR particles, we assume it to be the charge of proton in the following calculation for simplicity. $E$ is the energy of the GCR particles. $dN(E)/dE$ is the flux of the GCRs and a single power law spectrum ($\propto E^{- \gamma}$) with index $\gamma \approx 2.7$ is used in this work. $v(E)$ stands for the energy dependent streaming velocity of the GCRs, which can be estimated from the measurement of the anisotropy of GCRs. The integration is begun from 1 GeV and ended to 100 TeV as the contribution from higher energy is ignorable.

The velocity of the GCR streaming is estimated by simply assuming the anisotropy of the GCRs is due to the CG effect \citep{com35}
\begin{equation}
\frac{v(E)}{c} \approx \frac{1}{2+\gamma} \cdot \xi(E),
\end{equation}
where $c$ is the speed of light, and $\xi(E)$ is energy dependent anisotropy of GCRs \citep{par92}. According to the observations of various CR experiments, the amplitude of anisotropy increases with energy when the energy is below 1 TeV but nearly a constant when energy is between 1 and 100 TeV  \citep{ame05,gui07}. The anisotropy becomes to decrease when energy is higher than 100 TeV. For the purpose of an easy calculation, the energy dependent amplitude is parameterized by the following expression:
\begin{equation}
\xi(E) = \begin{cases}
a\times E^b  & \ (0.1 \ TeV<E<1 \ TeV) \\
a & \ (1 \ TeV\leq E<100 \ TeV) ,
\end{cases}
\end{equation}
where $a\sim 0.0005$, $b\sim 0.44$ are fitted from the observations. The energy $E$ is in a unit of TeV. For lack of the direct measurement, the amplitude of anisotropy below 0.1 TeV is estimated by the extrapolation of the above parameterization for energy between 0.1-1 TeV.

As a result, the current density in the solar vicinity induced by the GCR plasma with energy above 1 GeV is $\sim3.1\times 10^{-19} \ A m^{-2}$.

As shown in figure 1 (b), two kinds of Galactic electric currents are introduced:

	1) Inward currents along spiral arms

The spiral arm of the Galaxy or the magnetic field in disk can be described by a logarithmic spiral function \citep{hou09}
\begin{equation}
R = re^{\theta \text{cot}\varphi}.
\end{equation}
Here $(R,\theta)$ are the polar coordinates and $\varphi$ is the angle between the tangent and the radial line at the point $(R,\theta)$. The complementary angle of $\varphi$ is called the pitch angle, and the mean pitch angle of Galactic spiral arm is $12^{\circ}$ \citep{val95}. When $0<\theta<3\pi$ and $R<10 \ kpc$ the constant $r$ is  1.35.

We further assume that the inward current is mainly confined in the disk plane. It is approximated by an exponential decay term in the Z direction (i.e., the direction perpendicular to the Galactic plane),
\begin{equation}
j^{in}_{cr}(Z) = j_{0}^{in}e^{-|Z|/H_1},
\end{equation}
where $H_1$ is the characteristic decay length. $H_1=100 pc$, the half-thickness of the inner disk \citep{ptu09}, is adopted in this work. $j_{0}^{in}$ is the inward current density in Galactic plane (Z=0). Using Equation (1)-(3), we already know the current density in solar vicinity, i.e. $j^{in}_{cr}(Z=10pc)$, is $3.1\times 10^{-19} \ A m^{-2}$, so $j_{0}^{in}$ is $3.4\times 10^{-19} \ A m^{-2}$ according to Equation (5).

  2) Outward currents perpendicular to the galactic disk

Similarly, the outward currents to the NGP (for $Z>0$) and the SGP (for $Z<0$) are described by exponential functions in both the Z and R (radial) directions
\begin{equation}
j^{out}_{cr}(R,Z) = j_{0}^{out}e^{-|Z|/H_2}e^{-|R|/R_0}.
\end{equation}
We adopt $H_2=4 kpc$, and $R_0=4 kpc$, in order to match the model of the halo magnetic field described by \citet{sun10}. It is reasonable for such a value of $H_2$, due to the large uncertainty of the CR propagation halo size which could be larger than 10 kpc according to the present data \citep{put10}. The exponential term of the radial direction represents the feature that the density of CR sources is higher in the Galactic center region \citep{cas96}. $j_{\mathrm{0}}^{\mathrm{out}}$ is the outward current density in GC, which is $2.6\times 10^{-18} \ \mathrm{A m^{-2}}$ according to Equation (6), in the case that the current density in solar vicinity, $j^{\mathrm{out}}_{\mathrm{cr}}(R = 8.5 \mathrm{kpc}, Z = 10 \mathrm{pc})$ is $3.1\times 10^{-19} \ \mathrm{A m^{-2}}$ (obtained by Equations (1)-(3), as mentioned above).
\section{GALACTIC HALO MAGNETIC FIELD}

To calculate the possible contribution of GCR to the GMF in the halo, firstly, we need to consider the electric currents from both the relativistic CRs and the thermal plasma \citep{dol04}. The total current induced by these particles can be expressed as:
\begin{equation}
j = j_{cr}+j^{p}_{th}+j^{e}_{th}.
\end{equation}
Here $j_{cr}$ is the current density induced by relativistic GCR particles, $j^{p}_{th}$ and $j^{e}_{th}$ are the current densities induced by thermal protons and electrons respectively.

As described in \citep{dol04}, the diffusion coefficients for relativistic particles are not correlated with that for thermal particles. Thus, the two terms of the currents cannot cancel each other out, and should generate the magnetic field of the same order of magnitude. In this case, we can use the current generated by relativistic CR particles as an order of magnitude approximation. It should be noted that the total current is divergence-free due to the charge conservation.

According to \citep{dol04}, the evolution of the total GMF due to the electric currents of energetic particles can be solved by the following equation:
\begin{equation}
\frac{\partial B}{\partial t} = (\nu_{m} + \nu_{turb})\nabla ^{2} B + \nabla \times (u\times B) + \frac{4 \pi \nu_{m}}{c}\nabla \times j_{cr} .
\end{equation}
Here, $\nu_{m} = c^{2} /4 \pi \sigma$  is the local magnetic diffusion coefficient. $\nu_{m}$ varies in different phases of the ISM: $3 \times 10^{21} \ cm^{2}s^{-1}$ for cold $(T \sim 100K)$, $3 \times 10^{20} \ cm^{2}s^{-1}$ for warm $(T\sim 10^{4}K)$ and $5 \times 10^{17} \ cm^{2}s^{-1}$ for hot $(T > \sim 10^{5}K)$ phase \citep{dol04}, and the ratio of filling factor for these phases roughly is $0.5: 7.5:2$ \citep{fer01}. Therefore, the value averaged over the Galactic halo is $\nu_{m} \approx 3 \times 10^{20} \ cm^{2}s^{-1}$. $u$ is the regular component of the plasma velocity.

$\nu_{turb}$ is the turbulent diffusion coefficient. It can be estimated using the formula $\nu_{turb} \approx u'L/3$, where $L \approx 100  pc$  for typical component of the interstellar turbulence in the disk and $u' \approx 10 \ kms^{-1}$ for the characteristic turbulent velocity. It gives $\nu_{turb} \approx 10^{26} \ cm^{2}s^{-1}$ in the disk. To the main concern of this work, i.e., the Galactic halo, $\nu_{turb}$ has been estimated to be 30 times higher by \citep{ruz88}, i.e., $\nu_{turb} \approx 3 \times 10^{27} \ cm^{2}s^{-1}$.

With these parameters, the diffusion time for the magnetic field to fill the Galaxy, i.e. the time to reach the steady state, can be estimated by $\tau \approx R^{2}/(\nu_{turb} + \nu_{m}) \approx 10^{10} \ years$ for Galactic radius $R = 10 kpc$, which is in the same order as the age of the Galaxy. In other words, the current magnetic field of our galaxy is close to the stationary solution of the above equation, and the calculation of a steady magnetic field can be even simplified.

In \citep{dol04}, the above equation is obtained by applying an operation of $\nabla \times$ on both side of the following equation:
\begin{equation}
cE = - \frac{4 \pi \nu _{m}}{c}j_{cr} + (\nu_{m} + \nu_{turb})\nabla \times B - u\times B.
\end{equation}
Considering that for the stationary solution, and the electric field associated with a plasma moving in a magnetic field is given by $E = -u\times B/c$ in the limit of large electrical conductivity, which is satisfied in most astrophysical circumstances \citep{par79}, Equation (9) can be rewritten as
\begin{equation}
\nabla \times B = \frac{4 \pi}{c}\frac{\nu_{m}}{\nu_{m}+\nu_{turb}}j_{cr}.
\end{equation}
Comparing with the familiar Maxwell equation in the vacuum, we can see that the magnetic field in our case is equivalent to the solution for a largely suppressed electric current in the case of the vacuum.

The result of the calculated magnetic field is shown in Figure 2. We can see clearly two components of the magnetic field, a poloidal component and a toroidal component. The poloidal magnetic field is similar to a dipole pattern, with the direction from north to south, passing through the disk plane. The structure of the toroidal component is anti-symmetric, with a counter-clockwise direction in the north Galaxy and a clockwise direction in the south. Both structures are qualitatively consistent with the observations \citep{han97,han99}.

The radial profiles of the magnetic field at different Galactic heights (Z) are shown in Figure 3. We can see that the strength of the magnetic field decreases with the increase of Z. For the poloidal component, there is no observation yet to be compared directly. For the toroidal component, we give the parameterized description of observations, according to \citep{sun10}. Our theoretical calculation shows a rough agreement of the structures with the observations. The magnetic field strength in our calculation is approximately 2\% of the observation.

\section{DISCUSSION AND SUMMARY}

The observational anisotropy has a stable large-scale structure in the energy range from tens of GeV to hundreds of TeV. These structures imply three directions of the GCR streaming: inward and outward along the tangential direction of the spiral arm and the perpendicular direction to the Galactic pole. In this work, we extend the anisotropy observed locally to the whole Galaxy, which means the GCR streamings might exist at the global scale. A loop of the GCR streaming gives a new signature of the propagation of GCRs, and also provides contributions to the GMF.

Qualitatively, the GCR streamings inferred from the anisotropy of the GCRs can generate a large-scale magnetic field with poloidal and toroidal structures. These structures are consistent with the observations. This indicates that the extension of the local anisotropy to the whole Galaxy is probably reasonable. It will be helpful for further understanding of the origin of GCR anisotropy. Moreover, the anisotropy of the GCRs provides a new measure to study the Galactic electric current as well as a new window to understand the origin of the GMF.

This analysis also indicates that the magnetic field in the Galactic halo may be partially contributed by the electric current induced by GCRs. The extension of the local anisotropy to the whole Galaxy is too simple in this work and the uncertainties introduced by the parameters used in the calculation are quite large, so the quantitative result only can be regarded as an order-of-magnitude estimate.

\acknowledgments
We thank W. Y. Chen, S.F. Yang, and J. H. Xu for their questions, which stimulated us to conduct this work. We are grateful to C. Q. Shen and Y. G. Xie for reading the manuscript and providing comments on the writing of the Letter. We express our gratitude to Q. Yuan, P. F. Yin, J. Zhang and M. M. Kang for the beneficial discussions. This work was supported by the Ministry of Science and Technology of China, the Natural Sciences Foundation of China (No. 10725524, 11105156) and by the Chinese Academy of Sciences (Nos. KJCX2-YW-N13, GJHZ1004).

\clearpage

\begin{figure}
\includegraphics[height=5 cm]{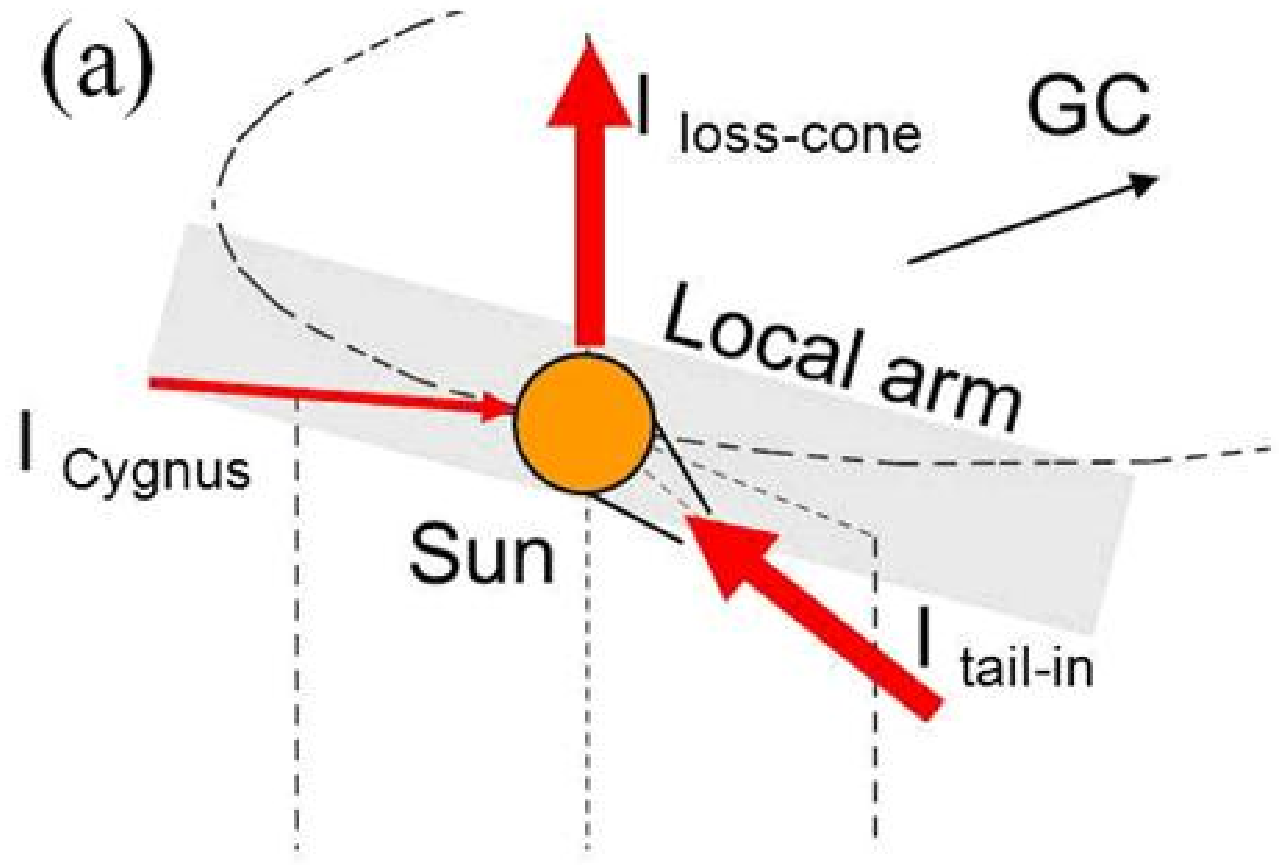} \\
\includegraphics[height=3 cm]{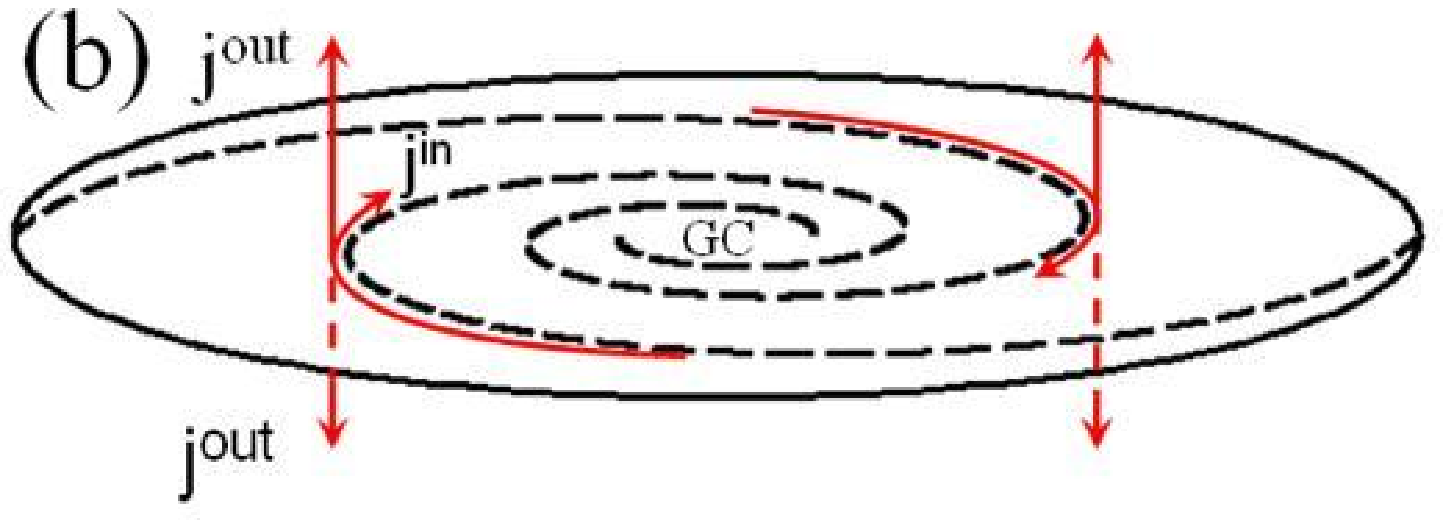}
\caption{(a) Schematic view of the anisotropy of the GCRs observed at the solar position. The three red arrows correspond to three anisotropy components, namely, the \textquotedblleft tail-in\textquotedblright \ excess, the \textquotedblleft loss-cone\textquotedblright \ deficit, and the Cygnus excess, respectively. The directions of their centers are noted as $I_{tail-in}$, $I_{loss-cone}$ and $I_{cygnus}$ respectively. $I_{tail-in}$ and $I_{cygnus}$ are approximately opposite along the local arm. $I_{loss-cone}$ is perpendicular to the Galactic plane. (b) Cartoon picture of the GCR streamings after the extension of the local anisotropy to the whole Galaxy. Inward streamings $(j^{in})$ are along the spiral arms and outward streamings $(j^{out})$ are perpendicular to the Galactic disk.}
\end{figure}

\clearpage

\begin{figure}
\includegraphics[height=5 cm]{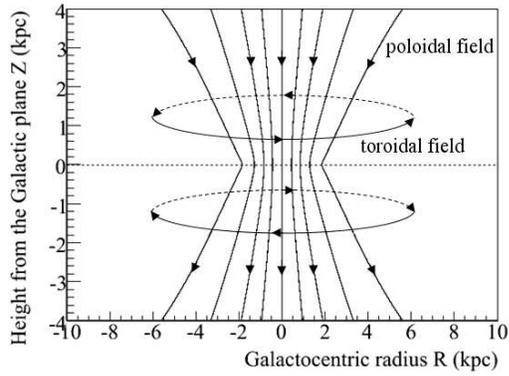}
\caption{Structure of the Galactic halo magnetic field contributed by the GCRs. The GMF in the halo is composed of a poloidal magnetic field and a toroidal magnetic field. The direction of the poloidal field is from Galactic north to south, passing through the disk plane. The toroidal field has reverse directions below and above the Galactic plane. The arrows indicate the directions of the magnetic fields.}
\end{figure}

\clearpage

\begin{figure}
\includegraphics[height=5 cm]{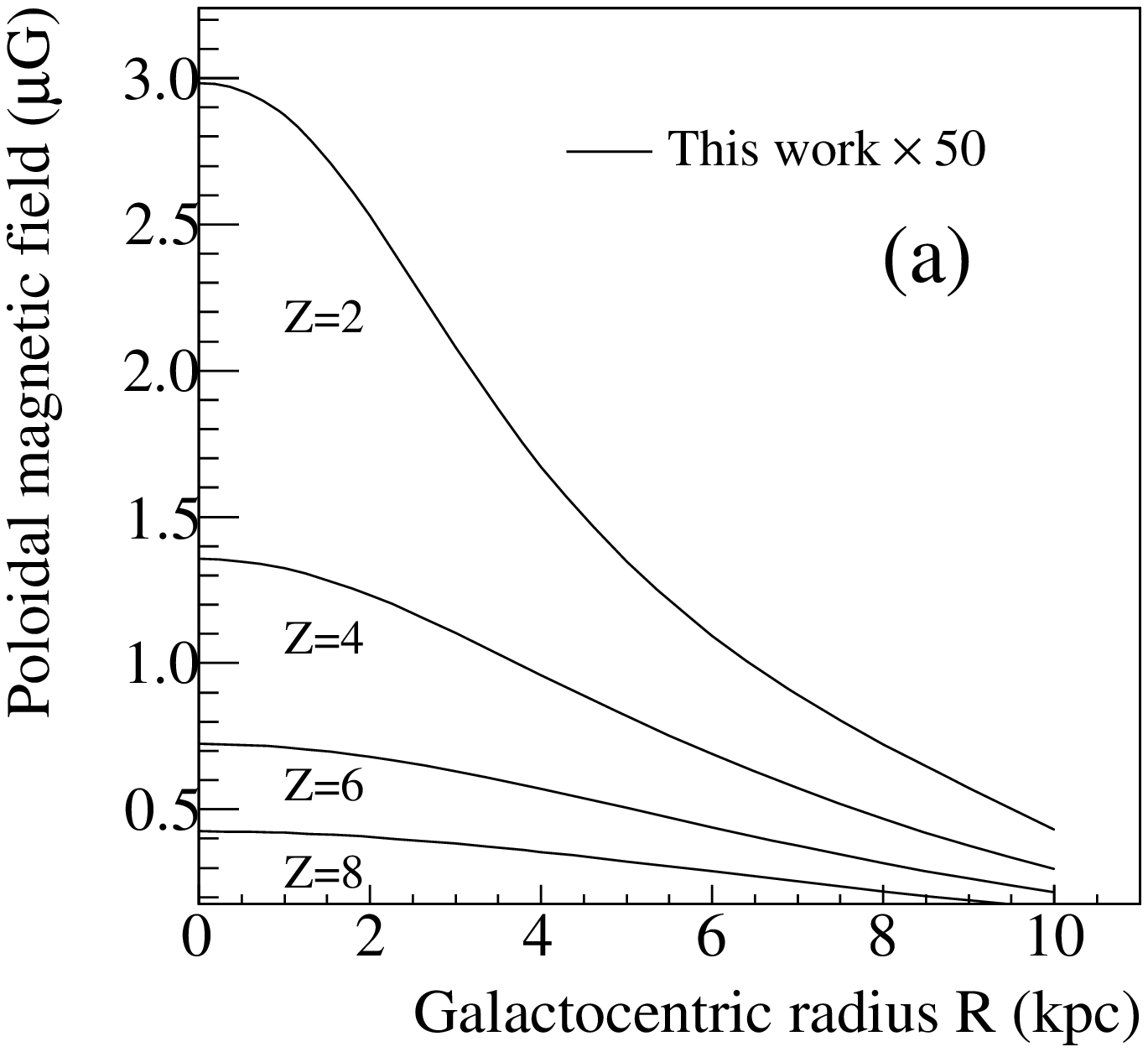}
\includegraphics[height=5 cm]{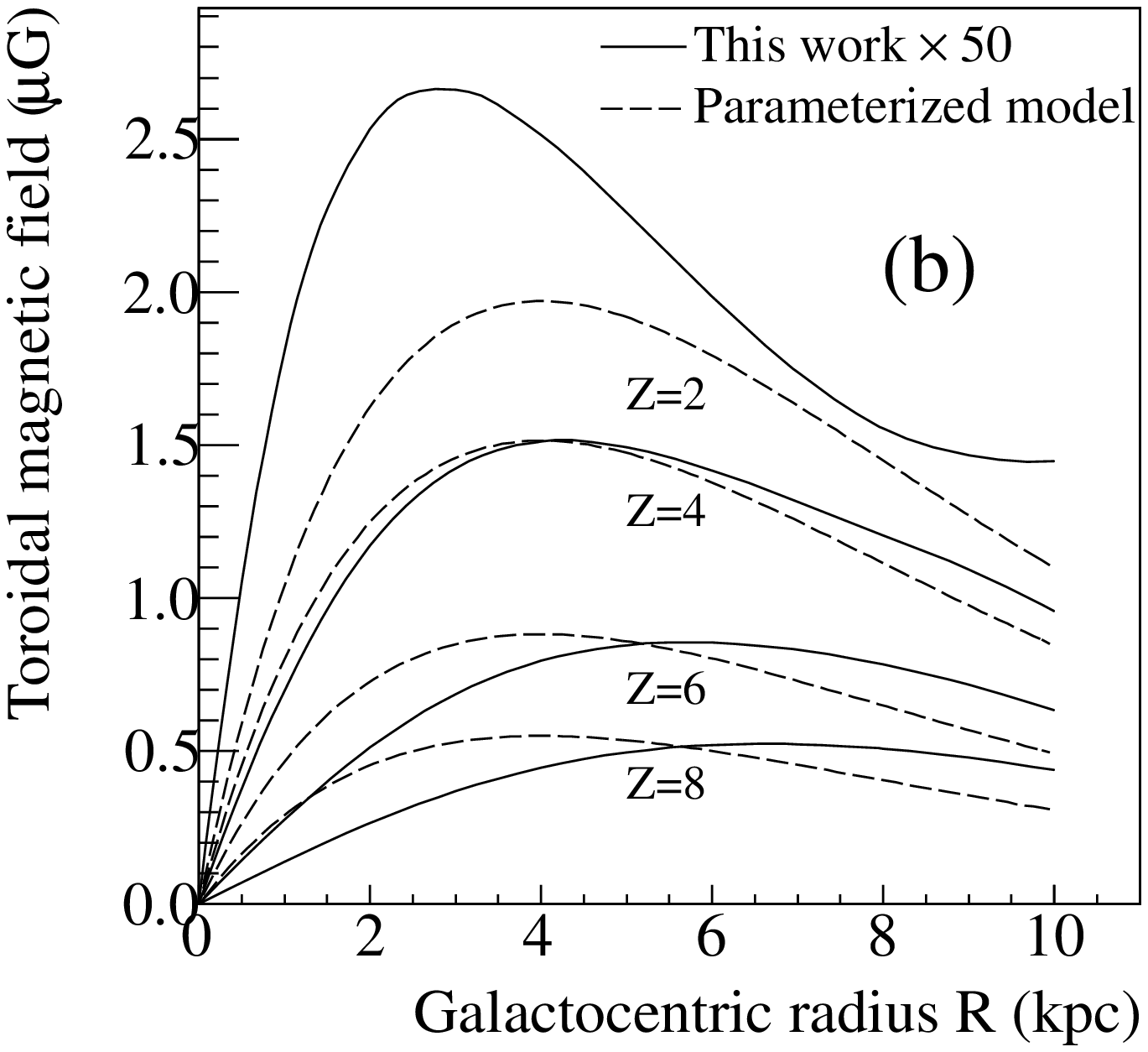}
\caption{Distribution of the poloidal (a) and toroidal (b) magnetic field strength. The solid line is the strength of the magnetic field calculated by our model as a function of the Galactocentric radius (R) at a given height (Z). The dashed lines in (b) present the strength distribution of the toroidal magnetic field derived from the parameterized model \citep{sun10}. The strength of the magnetic field (solid line) is the magnetic field calculated by our model multiplied by 50 for the convenience to compare the distribution behavior. The solid lines and the dashed lines from top to bottom correspond to the height at 2, 4, 6, and 8 kpc, respectively. }
\end{figure}

\clearpage

\end{document}